\documentclass[a4paper,USenglish]{lipics-v2016}
 
\usepackage{microtype}

\theoremstyle{plain}
\newtheorem{my_theorem}{Theorem}

\theoremstyle{definition}
\newtheorem{my_definition}{Definition}
\newtheorem{my_example}{Example}

\theoremstyle{remark}


\newtheorem{proposition}{Proposition}

\usepackage{xcolor}

\usepackage{hyperref}
\hypersetup{backref=true,       
    pagebackref=true,               
    hyperindex=true,                
    colorlinks=true,                
    breaklinks=true,                
    urlcolor= black,                
    linkcolor= blue,                
    bookmarks=true,                 
    bookmarksopen=false,
    filecolor=black,
    citecolor=blue,
    linkbordercolor=blue
}
\AtBeginDocument{\hypersetup{pdfborder={0 0 1}}}
\usepackage{algpseudocode}
\algdef{SE}[DOWHILE]{Do}{doWhile}{\algorithmicdo}[1]{\algorithmicwhile\ #1}%

\algnewcommand{\IIf}[1]{\State\algorithmicif\ #1\ \algorithmicthen}
\algnewcommand{\EndIIf}{\unskip\ \algorithmicend\ \algorithmicif}
\algdef{SE}[DOWHILE]{Do}{doWhile}{\algorithmicdo}[1]{\algorithmicwhile\ #1}%

\usepackage{amssymb}

\title{Graph Pattern Matching Preserving Label-Repetition Constraints}
\titlerunning{Graph Pattern Matching Preserving Label-Repetition Constraints} 

\author[1]{Houari Mahfoud}
\affil[1]{Abou-Bekr Belkaid University \& LRIT Laboratory, Tlemcen, Algeria\\
  \texttt{houari.mahfoud@gmail.com}}
\authorrunning{H. Mahfoud} 


\begin{document}

\maketitle

\begin{abstract}
Graph pattern matching is a routine process for a wide variety of applications such as social network analysis.
It is typically defined in terms of subgraph isomorphism which is \textsc{NP-Complete}. 
To lower its complexity, many extensions of graph simulation have been proposed which 
focus on some topological constraints of pattern graphs that can be preserved in polynomial-time over data graphs. 
We discuss in this paper the satisfaction of a new topological constraint, called \emph{Label-Repetition constraint}. 
To the best of our knowledge, existing polynomial approaches fail to preserve this constraint, and moreover, one can adopt only subgraph 
isomorphism for this end which is cost-prohibitive. 
We present first a necessary and sufficient condition that a data subgraph must satisfy to preserve the 
\emph{Label-Repetition constraints} of the pattern graph. 
Furthermore, we define matching based on a notion of \emph{triple simulation}, an extension of graph simulation by considering the 
new topological constraint. We show that with this extension, graph pattern matching can be performed in polynomial-time, by providing such an 
algorithm. Our algorithm is sub-quadratic in the size of data graphs only, and quartic in general. We show that our results can be combined with 
orthogonal approaches for more expressive graph pattern matching.
\end{abstract}

\section{Introduction}\label{sec:Introduction}
Modeling data with graphs is one of the most active topics in the database community these days. This model has recently gained 
wide applicability in numerous domains that find the relational model too restrictive, such as social networks \cite{GPM_For_Social_Network}, 
biological networks, Semantic Web, crime detection networks and many others. Indeed, it is less complex and also most natural for users to reason 
about an increasing number of popular datasets, such as the underlying networks of Twitter, Facebook, or LinkedIn, within a graph paradigm. 
In emerging applications such as social networks, edges of data graphs (resp. pattern graphs) can be typed \cite{Adding_Reg_Exp_Fan_11} to 
denote various relationships such as marriage, friendship, recommendation, co-membership, etc. Moreover, pattern graphs can define 
multi-labeled vertices \cite{Multi-Labeled-Nodes} to look, e.g., for persons with different possible profiles.


Given a data graph $G$ and a pattern graph $Q$, the problem of \emph{graph pattern matching} is to find all subgraphs of $G$ that 
satisfy both the labeling properties and topological constraints carried by $Q$. Matching here is expressed in terms of subgraph isomorphism 
which consists to find all subgraphs of $G$ that are \emph{isomorphic} to $Q$. Graph pattern matching via subgraph isomorphism is an 
\textsc{NP-Complete} problem as there are possibly an exponential number of subgraphs in $G$ that match $Q$. 
To tackle this \textsc{NP-Completeness}, graph simulation \cite{Milner1989} has been adopted for graph pattern matching \cite{GraphSimulation}
to preserve child-relationships only. Unlike subgraph isomorphism which requires a \emph{bijective} mapping function from pattern nodes to 
data nodes, graph simulation is defined by a simple \emph{binary} relation which can be computed in quadratic time. 
A cubic-time extension of graph simulation, called \emph{strong simulation}, has been proposed \cite{Fan14} by enforcing two additional conditions: 
\emph{duality} to preserve child and parent relationships of the pattern graph; and \emph{locality} to overcome excessive matching 
by considering only subgraphs that have radius bounded by the diameter of the pattern graph.


Nonetheless, the polynomial-time complexity comes at a price: the result of strong simulation may contain incorrect matches as shown below.

\begin{figure}[t]
\centering
   \includegraphics[width=280px,height=140px]{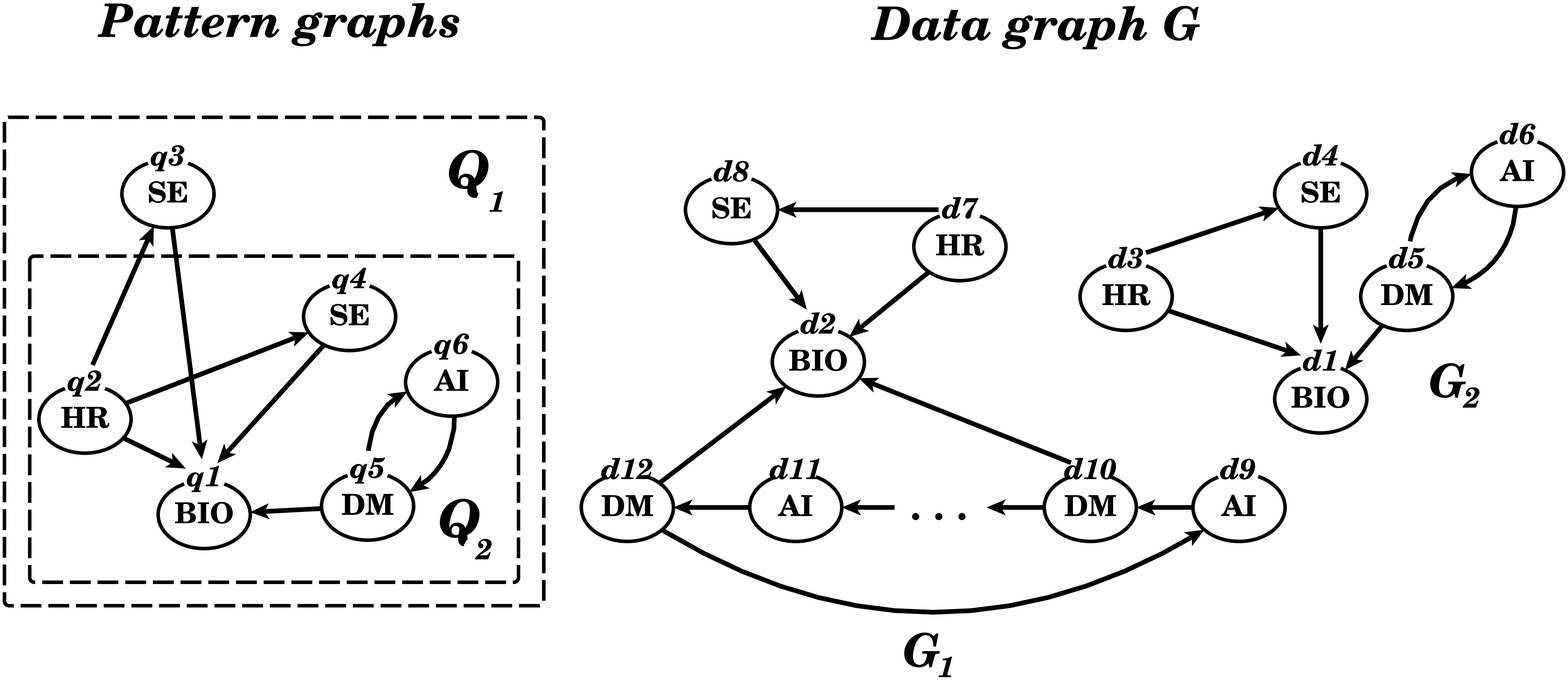}
   \caption{Querying a recommendation network.}
   \label{Introductory_Example_Figure}
\end{figure}

\begin{my_example}\label{Introductory_Example}
Consider the real-life example taken from \cite{Fan14} with minor modification. 
A headhunter (\textbf{HR}) wants to find a biologist (\textbf{BIO}) to help a group of software engineers (\textbf{SE}) analyze genetic data. 
To do this, she uses the network $G$ depicted in Fig. \ref{Introductory_Example_Figure}. In $G$, nodes denote persons 
with different profiles, and edges indicate recommendations between these persons. The cycle between the nodes $d_{9}$ and $d_{12}$ contains many 
\textbf{DM} (data mining specialist) that are all connected to the \textbf{BIO} represented by the node $d_2$. 
The biologist \textbf{BIO} to find is specified with the pattern graph $Q_1$ of Fig. \ref{Introductory_Example_Figure}. 
Intuitively, the \textbf{BIO} has to be recommended by: (a) an \textbf{HR} person since the headhunter trusts the judgment
of a person with the same occupation; (b) at least two \textbf{SE} that are recommended by the 
same \textbf{HR} person (to increase incredibility), that is, the \textbf{BIO} has a strong experience by working with different \textbf{SEs}; 
and (c) a \textbf{DM}, as data mining techniques are required for the job. Moreover, there is an artificial intelligence expert (\textbf{AI}) who 
recommends the \textbf{DM} and is recommended by a \textbf{DM}. 

When strong simulation is adopted, the subgraph $G_2$ of $G$ is returned as the only match of $Q_1$ in $G$. However, 
the \textbf{BIO} of this match, represented by the node $d_1$, is recommended by only one \textbf{SE}, which is incorrect w.r.t $Q_1$. 
To make search less restrictive, one can look for a \textbf{BIO} with the same constraints specified by $Q_1$ excepting that this \textbf{BIO} 
can be recommended by only one \textbf{SE}. This search is specified by the pattern graph $Q_2$ of the same figure. 
In this case, strong simulation returns $G_2$ as the only match of $Q_2$ in $G$, which is a correct. Notice however that strong simulation 
does not make difference between $Q_1$ and $Q_2$ since the two pattern graphs are matched over $G$ to the same match result.
\end{my_example}

The pattern graph $Q_1$ illustrates a new kind of topology that we call \emph{Label-Repetition} (\emph{LR}) constraint. 
Graph simulation \cite{GraphSimulation} and its counterparts \cite{GPM_From_Intractable_To_Polynomial_Time, Fan14} fail to preserve this constraint. 
One can adopt subgraph isomorphism to preserve \emph{LR} constraints during graph pattern matching. The challenge is that 
subgraph isomorphism is \textsc{NP-Complete} and real-life data graphs are often big, e.g., 
the social graph of Facebook has billions of nodes and trillions of edges \cite{FacebookStatistic}. This motivates us to study an extension of 
graph simulation in order to preserve \emph{LR} constraints in polynomial-time.\newline

\noindent\textbf{Contributions \& Road-map.} Our main contributions are as follows:\footnote{The proofs are given in Appendix.}
\textbf{(1)} We introduce a new extension of graph simulation, called \textit{triple simulation}, to preserve \emph{LR} 
constraints (Section \ref{sec:TripleSimulation}). \textbf{(2)} We define a necessary and sufficient condition that characterizes the satisfaction 
of \emph{LR} constraints and we compute its time complexity (Section \ref{sec:Satisfy_LO_Properties}). \textbf{(3)} We develop a graph 
pattern matching algorithm which requires a polynomial-time to preserve \emph{Child} and \emph{Parent} relationships, as well as \emph{LR} 
constraints (Section \ref{section:Algorithm_for_TripleSimulation}). Finally, we show how to improve the quality of our match results by using the 
notion of locality (Section \ref{section:TSim_With_Locality}).\newline

\noindent\textbf{Related work.} We categorize related work as follows.

\noindent\textit{\underline{Polynomial-time graph pattern matching:}} Traditional matching is by subgraph isomorphism, which is 
NP-Complete \cite{Subgraph_Isomorphism_NP_Completness} and found often too restrictive to capture sensible 
matches \cite{GPM_From_Intractable_To_Polynomial_Time}. To loosen the restriction, one direction is to adopt 
\emph{graph simulation} \cite{Milner1989}. Matching based on graph simulation \cite{GraphSimulation} preserves only child relationships of the pattern graphs, 
which makes it useful for some applications like Web sites classification \cite{GraphSimulation_Use_Case}. In other applications however, 
e.g. social network analysis, the result of such matching may have a structure drastically different from that of the pattern graph, 
and often 
very large to analysis and understand. To handle this, \emph{strong simulation} is proposed \cite{Fan14} to capture child and parent relationships 
(notion of \emph{duality}), and to make match results bounded by the diameter of the underlying pattern graph (notion of \emph{locality}). 
This approach has proven efficient since it is in PTIME. However, it can not match correctly pattern graphs with \emph{LR} constraints.

\noindent\textit{\underline{Quantified pattern graphs:}} Closer to our work is \cite{Counting_Quantifiers_Fan_16} that introduces 
\emph{quantified pattern graphs} (\textbf{QGPs}), an extension of pattern graphs by supporting simple counting
quantifiers on edges. A \textbf{QGP} naturally expresses numeric and ratio aggregates, and negation besides existential and universal 
quantification. Notice that any ratio aggregate can be translated into numeric aggregate. 
Quantified matching is NP-Complete in the absence of negation and DP-Complete for general \textbf{QGPs}. 
As shown in the Appendix \ref{appendix_TSim_CQs}, any \textbf{QGP} with numeric aggregates 
can be translated into a simple pattern graph with only \emph{LR} constraints. This translation allows to preserve numeric and ratio aggregates on 
edges in polynomial-time, contrary to the prohibitive-cost found by the authors \cite{Counting_Quantifiers_Fan_16}. 
Furthermore, we think that matching over pattern graphs with negation and universal quantifications on edges can be done in PTIME if treated as an 
extension of graph simulation (one of our future directions).

\section{Background}\label{sec:Background}
We give basic notions of graphs and then we review some graph pattern matching approaches.

\textbf{Graphs.} A \emph{directed graph} (or simply a \emph{graph}) is defined with $G$($V,E,\lambda$) where: 1) $V$ is a finite set of nodes; 
2) $E\subseteq V\times V$ is a finite set of edges in which $(u,u^{'})$ denotes an edge from nodes $u$ to $u^{'}$; and 3) $\lambda$ is a labeling 
function that maps each node $u\in V$ to a label $\lambda(u)$ in a set $\sum(G)$ of labels. We simply denote $G$ as $(V,E)$ when it is clear from 
the context.

\noindent In this paper, both data graphs and pattern graphs are specified with the previous graph structure.
Moreover, we assume that pattern graphs are connected, as a common practice.

\textbf{Distance and diameter \cite{Fan14}.} The \emph{distance} from nodes $n$ to $n^{'}$ in a graph $G$,
denoted by \emph{dist}($n,n^{'}$), is the length of the shortest undirected path from $n$ to $n^{'}$ in $G$. 
The \emph{diameter} of a connected graph $G$, denoted by $d_G$, is the longest shortest distance of all pairs of nodes in $G$, 
that is, $d_G$ = \emph{max}(\emph{dis}($n$, $n^{'}$)) for all nodes $n$, $n^{'}$ in $G$.

\textbf{Graph pattern matching.} A data graph $G$($V,E,\lambda$) may match a pattern graph $Q$($V_Q,E_Q,\lambda_Q$) via different methods.

\noindent\textit{\underline{A) Subgraph isomorphism:}} A subgraph $G_s$($V_s,E_s,\lambda_s$) of $G$ matches $Q$ via \emph{subgraph isomorphism}, 
denoted $G_s\prec_{iso} Q$, if there exists a \emph{bijective function} $f$:$V_Q\rightarrow V_s$ such that: 
1) for each node $n\in V_Q$, $\lambda_Q(n)=\lambda_s(f(n))$; and 2) for each edge $(n, n^{'})\in E_Q$, 
there exists an edge $(f(n), f(n^{'}))\in E_s$.

\noindent\textit{\underline{B) Graph simulation:}} $G$ matches $Q$ via \emph{graph simulation} \cite{GraphSimulation}, 
denoted $Q\prec G$, if there exists a \emph{binary match relation} $S\subseteq V_Q\times V$ such that: 
\begin{enumerate}
 \item For each $(u, v)\in S$, $\lambda_Q(u)=\lambda(v)$; and
 \item For each node $u\in V_Q$, there exists a node $v\in V$ such that: a) $(u,v)\in S$; and b) for each edge $(u,u^{'})\in E_Q$, there exists an 
edge $(v,v^{'})\in E$ with $(u^{'},v^{'})\in S$.
\end{enumerate}
Intuitively, graph simulation preserves only child relationships of the pattern graph.

\noindent\textit{\underline{C) Dual simulation:}} $G$ matches $Q$ via \emph{dual simulation} \cite{Fan14}, 
denoted $Q\prec_D G$, if there exists a \emph{binary match relation} $S_D\subseteq V_Q\times V$ such that: 
\begin{enumerate}
 \item For each $(u, v)\in S_D$, $\lambda_Q(u)=\lambda(v)$; and
 \item For each node $u\in V_Q$, there exists a node $v\in V$ such that: a) $(u,v)\in S_D$; b) for each edge $(u,u^{'})\in E_Q$, there exists an 
edge $(v,v^{'})\in E$ with $(u^{'},v^{'})\in S_D$; and moreover c) for each edge $(u^{'},u)\in E_Q$, there exists an 
edge $(v^{'},v)\in E$ with $(u^{'},v^{'})\in S_D$.
\end{enumerate}
Remark that dual simulation enhances graph simulation by imposing the condition (c) in order to preserve both child and parent relationships. 
As mentioned in \cite{Fan14}, the graph pattern matching via graph simulation (resp. dual simulation) is to find the the \emph{maximum} match 
relation $S$ (resp. $S_D$). 
Ma et al. \cite{Fan14} show that graph/dual simulation may do excessive matching of pattern graphs which makes the graph result 
very large and difficult to understand and analysis. For this reason, they propose \emph{strong simulation}, an extension of dual simulation by 
imposing the notion of \emph{locality}. This notion requires that each subgraph of the final match result must have a radius bounded by the 
diameter of the pattern graph.

\noindent\textit{\underline{D) Strong simulation:}} $G$ matches $Q$ via \emph{strong simulation}, denoted $Q\prec^{L}_{D} G$, if there exists a node 
$v\in V$ and a subgraph $G_s$ of $G$ centered at $v$ such that:
\begin{enumerate}
 \item The radius of $G_s$ is bounded by $d_Q$, i.e., for each node $v^{'}$ in $G_s$, \emph{dist}($v,v^{'}$)$\leq d_Q$;
 \item $Q\prec_D G_s$ with the maximum match relation $S_D$.
\end{enumerate}
Informally, rather than matching the whole data graph $G$ over $Q$ we extract, for each node $n\in V$, a subgraph $G_s$ of $G$ centered at $n$ and 
which has a radius equals to $d_Q$. Then, we match $G_s$ over $Q$ via dual simulation. In this way, the match result will be composed of subgraphs 
of reasonable size that satisfy both child and parent relationships of $Q$.

\textbf{Match results.} \textit{\textbf{A)}} When $Q\prec_{iso} G$ then the \emph{match result} $\mathcal{M}_{iso}(Q,G)$ 
is the set of all subgraphs of $G$ that are isomorphic to $Q$. \textit{\textbf{B)}} When $Q\prec G$ with the maximum match relation $S$ then the 
\emph{match result} $\mathcal{M}(Q,G)$ w.r.t $S$ is each subgraph $G$($V_s,E_s$) of $G$ in which: 1) a node $n\in V_s$ iff it is in $S$; 
and 2) an edge $(v,v^{'})\in E_s$ iff there exists an edge $(u,u^{'})\in E_Q$ with 
$(u,v)\in S$ and $(u^{'},v^{'})\in S$. \textit{\textbf{C)}} When $Q\prec_D G$ then the \emph{match result} $\mathcal{M}_D(Q,G)$ 
is defined similarly to graph simulation but w.r.t the maximum match relation $S_D$. 
\textit{\textbf{D)}} When $Q\prec_{D}^{L} G$ then the \emph{match result} $\mathcal{M}^{L}_{D}(Q,G)$ is defined with 
$\bigcup_{i}\mathcal{M}_{D}(Q,G_i)$ where each $G_i$ is a subgraph of $G$ that satisfies the conditions of strong simulation.

\textbf{Potential matches.} Given a data graph $G$($V,E,\lambda$) and a pattern graph $Q$($V_Q,E_Q,\lambda_Q$). For any node $u\in V_Q$, 
we call \emph{potential match} each node $v\in V$ that has the same label as $u$ (i.e. $\lambda_Q(u)=\lambda(v)$). 
Moreover, \textsc{sim}($u$) refers to the set of all potential matches of $u$ in $G$.

\begin{my_example}
Consider the data graph $G$ and the pattern graph $Q_2$ of Fig. \ref{Introductory_Example_Figure}. With dual simulation, both $G_1$ 
and $G_2$ are found as matches of $Q_2$ in $G$. Remark that the cycle of two nodes \textbf{AI} and \textbf{DM} in $Q_2$ is matched 
with the long cycle $d_9\rightarrow\dots\rightarrow d_{12}\rightarrow d_9$ in $G_2$, which may be hard to analysis. With the notion of locality, 
strong simulation returns $G_1$ as the only match of $G$ over $Q_2$ and ignores $G_2$ since it represents an excessive matching.
\end{my_example}

\section{Triple Simulation}\label{sec:TripleSimulation}
We start first by presenting a new topological constraint that one would like to preserve during graph pattern matching. We then define a new 
extension of graph simulation by imposing this constraint. 
We compare our extension with only strong simulation \cite{Fan14} since this is the more expressive graph pattern matching 
approach that requires a polynomial-time. Notice that another polynomial-time approach exists \cite{GPM_From_Intractable_To_Polynomial_Time}, 
called \emph{bounded simulation}, which imposes constraints on edges. However, our extension concerns nodes constraints.

Given a data graph $G$ and consider the pattern graphs $Q_1=a\rightarrow b$ and $Q_2=b\leftarrow a\rightarrow b$. It is obvious that these two 
patterns are not equivalent: $Q_1$ requires that each node $v$ in $G$ that matches $a$ must have \emph{\textbf{at least one}} child node labeled 
with $b$, however, $Q_2$ requires that $v$ must have \emph{\textbf{at least two}} child nodes labeled with $b$. Strong simulation fails to make 
this difference and considers $Q_1$ and $Q_2$ as equivalent patterns (as illustrated by Example \ref{Introductory_Example}).


\begin{my_definition}\label{definition:LR_Constraints}
Given a data graph $G$($V,E$) and a pattern graph $Q$($V_Q,E_Q$).
A \emph{Label-Repetition} (\textbf{\emph{LR}}) constraint defined over a node $u\in V_Q$ with label $l$ specifies that: 
\textbf{\textit{1)}} there is a maximum subset $C_u=\{u_1,\dots,u_K\}$ ($K\geq 2$) of children (resp. parents) of $u$ that are all 
labeled with $l$; and \textbf{\textit{2)}} any match $v$ of $u$ in $G$ must have a subset $C_v=\{v_1,\dots,v_K\}$ of children (resp. parents) 
ordered in such a way that allows to match each $v_{i}$ to a child $u_{i}$ of $u$.
\end{my_definition}

Intuitively, a \emph{LR} constraint concerns a repetition of some label either among children or among parents of some node in $Q$. If children 
(resp. parents) of each node in $Q$ have distinct labels, then $Q$ is defined with only child and parent relationships and, thus, can be matched 
correctly via strong simulation. The limitation of this latter is observed when some children (resp. parents) of the same node are defined with 
the same label.

\begin{my_example}
Consider the pattern graph $Q_1$ of Fig. \ref{Introductory_Example_Figure}. There is an \emph{LR} constraint defined over the node $q_2$ with label 
\textbf{SE}. It specifies that each node of the data graph that matches $q_2$ must have at least two children labeled 
\textbf{SE} such that one of them matches the node $q_3$ and the other one matches the node $q_4$. 
\end{my_example}

We propose next a new extension of graph simulation in order to satisfy \emph{LR} constraints.

\begin{my_definition}\label{Definition_Of_TripleSimulation}
A data graph $G(V,E,\lambda)$ matches a pattern graph $Q(V_Q,E_Q,\lambda_Q)$ via \emph{triple simulation}, 
denoted by $Q \prec_T G$, if there exists a binary match relation $S_{T}\subseteq V_Q \times V$ s.t.:

\begin{enumerate}
    \item For each $(u,v)\in S_{T}$, $\lambda_Q(u)=\lambda(v)$.
    \item For each $u\in V_Q$ there exists $(u,v)\in S_{T}$.
    \item For each $(u,v)\in S_{T}$ and for all edges $(u, u_{1}),...,(u, u_{n})\in E_Q$, there exists 
    \emph{\textbf{at least $n$ distinct children}} $v_{1},...,v_{n}$ of $v$ in $G$ such that: $(u_{1}, v_{1}),...,(u_{n}, v_{n})\in S_{T}$.
    \item For each $(u,v)\in S_{T}$ and for all edges $(u_{1},u),...,(u_{n},u)\in E_Q$, there exists 
    \emph{\textbf{at least $n$ distinct parents}} $v_{1},...,v_{n}$ of $v$ in $G$ such that: $(u_{1}, v_{1}),...,(u_{n}, v_{n})\in S_{T}$.
\end{enumerate}
\noindent $\mathcal{M}_{T}(Q,G)$ is the match result that corresponds to the maximum match relation $S_{T}$\footnote{This match result can be defined similarly to graph (dual) simulation.}.
\end{my_definition}

Intuitively, if a node $u$ in $Q$ has $n$ children (resp. parents) then each match $v$ of $u$ in $G$ must have at least $n$ distinct 
children (resp. parents) such that we can match, w.r.t some order, each child (resp. parent) of $v$ to only one child (resp. parent) of $u$. 
This new restriction imposed by conditions (3) and (4) prevents matching of distinct children (resp. parents) of some 
node $u$ in $Q$ to the same node in $G$, as may be done by strong simulation. 
Notice that triple simulation preserves also child and parent relationships and not only \emph{LR} constraints.

\begin{my_example}
Consider the data graph $G$ and the pattern graphs $Q_1$ and $Q_2$ of Fig. \ref{Introductory_Example_Figure}.
The node $q_1$ with label \textbf{BIO} in $Q_1$ has two parents, $q_3$ and $q_4$, that have the same label \textbf{SE}. 
Remark that $d_1$ and $d_2$ are potential matches of $q_1$ in $G$. According to triple simulation, 
$d_1$ (resp. $d_2$) must have at least two distinct parents s.t. one can match $q_3$ and the other one can match $q_4$. 
This is not the case since $d_1$ (resp. $d_2$) has only one parent labeled \textbf{SE}. 
Thus, we can conclude that no subgraph in $G$ satisfies the \emph{LR} constraint of $Q_1$, and then, $\mathcal{M}_{T}(Q_1,G)=\emptyset$.
When triple simulation is adopted for $Q_2$ over the subgraph $G_2$, we obtain the following maximum match relation:
$S_{T}=\{(q_1,d_1),(q_2,d_3),(q_4,d_4),(q_5,d_5),(q_6,d_6)\}$. The match result that corresponds to $S_{T}$ is the whole subgraph $G_2$, which is 
correct.
\end{my_example}

We use \emph{\textbf{CPL} relationships} to refer to \emph{\textbf{C}hild} and 
\emph{\textbf{P}arent} relationships (called \emph{duality} properties), as well as relationships based on \emph{\textbf{L}R} constraints. 
Our motivation is to popose a graph pattern matching algorithm that preserves \emph{CPL} relationships in polynomial-time.

\begin{figure}[t]
\centering
   \includegraphics[width=\linewidth]{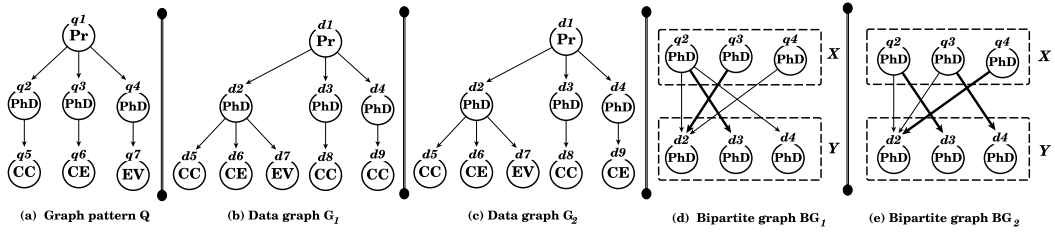}
   \caption{Problem of preserving \emph{LR} constraints.}
   \label{LO_Checking_Figure}
\end{figure}

\section{Satisfy \emph{LR} Constraints}\label{sec:Satisfy_LO_Properties}
We first present the problem of satisfying \emph{LR} constraints and show that a naive approach may lead for exponential cost. Next, we define a 
condition that is necessary and sufficient for the satisfaction of \emph{LR} constraints and which can be checked in polynomial-time.

\begin{my_example}\label{example:LO_Properties_Satisfaction}
Consider the graphs depicted in Fig. \ref{LO_Checking_Figure}. 
The pattern graph $Q$ looks for each professor (\textbf{Pr}) which has supervised at least three \textbf{PhD} thesis in topics related 
respectively to \emph{Cloud Computing} (\textbf{CC}), \emph{Collaborative Editing} (\textbf{CE}) and \emph{Electronic Vote} (\textbf{EV}).
The node $d_1$ in $G_1$ is a potential match of $q_1$. To satisfy the condition (3) of triple simulation 
(Definition \ref{Definition_Of_TripleSimulation}), $d_1$ must have at least three child nodes which is the case, and there must be some order that 
allows to match each child of $d_1$ to a child of $q_1$. 
However remark that: if we match $q_2$ with $d_2$ then we can not have match neither for $q_3$ nor for 
$q_4$; and moreover, if we match $q_2$ with $d_3$ then we can match either $q_3$ with $d_2$ or $q_4$ with $d_2$. Clearly, there is no order over 
the children $d_2, d_3, d_4$ of $d_1$ that allows to match all the children $q_2, q_3, q_4$ of $q_1$ in $Q$. 
Therefore, the data graph $G_1$ does not satisfy the \emph{LR} constraint of $Q$. 
On the other side, the data graph $G_2$ match correctly $Q$: see that there is an order that allows to match each child of $d_1$ to a child of 
$q_1$, i.e., $q_2, q_3, q_4$ can be matched respectively with $d_3, d_4, d_2$. Thus, the \emph{LR} constraint of $Q$ is satisfied over $G_2$.
\end{my_example}

Given the aboves, one can think that checking \emph{LR} constraints may lead to exponential cost (since we must consider all 
orders over some data nodes). However, we show later that this process can be done in polynomial-time.

\begin{my_definition}\label{definition:Bipartite_Graph_Inspecting_LO}
Given a data graph $G$($V,E$) and a pattern graph $Q$($V_Q,E_Q$). 
Consider all the \emph{LR} constraints defined over children (resp. parents) of some node $u\in V_Q$, and let $v\in V$ be a potential 
match of $u$. The bipartite graph $BG$($X\cup Y, E$) that \emph{inspects} these \emph{LR} constraints w.r.t $v$ is defined as follows:

\begin{itemize}
 \item $X\subseteq V_Q$ contains each child (resp. parent) of $u$ that is concerned by an \emph{LR} constraint.
 \item $Y\subseteq V$ contains each child (resp. parent) of $v$ that (potentially) matches some node in $X$.
 \item $(u^{'},v^{'})\in E$ if $u^{'}\in X$ is (potentially) matched with $v^{'}\in Y$.
\end{itemize}

\noindent A \emph{complete matching} over $BG$ is a maximum matching \cite{Cormen} that covers 
each node in $X$ \footnote{It is also called \emph{X-saturating matching}.}.
\end{my_definition}

Consider only the \emph{LR} constraints defined over children of $u$. The set $X$ of the bipartite graph $BG$ contains all children of $u$ that are 
concerned by some \emph{LR} constraint, and the set $Y$ contains each child of $v$ that (potentially) matches some child $u^{'}$ of $u$, 
provided that $u^{'}$ is concerned by an \emph{LR} constraint (i.e. $u^{'}\in X$). Moreover, an edge in $E\subseteq X\times Y$ denotes some child 
of $u$ in $X$ that can be (potentially) matched with some child of $v$ in $Y$.
For \emph{LR} constraints defined over parents of $u$, the bipartite graph that inspects them is defined in the same manner (i.e. $X$ is a subset 
of parents of $u$, and $Y$ is a subset of parents of $v$).

\begin{my_example}
Consider the pattern graph $Q$ and data graphs $G_1$ and $G_2$ depicted in Fig. \ref{LO_Checking_Figure}. 
Recall that there is an \emph{LR} constraint defined over the children of the node $q_1$ in $Q$. The bipartite graph $BG_1$ 
that inspects this \emph{LR} constraint, w.r.t the potential match $d_1$ of $q_1$ in $G_1$, is depicted in Fig. \ref{LO_Checking_Figure} (d). 
Moreover, w.r.t the potential match $d_1$ of $q_1$ in $G_2$, the corresponding bipartite graph $BG_2$ is given in 
Fig. \ref{LO_Checking_Figure} (e).
\end{my_example}


The next theorem states our main contribution which is a \emph{necessary and sufficient} condition to satisfy \emph{LR} constraints.

\begin{my_theorem}\label{theorem:NS_Condition_For_LO}
Given a data graph $G$($V,E$), a pattern graph $Q$($V_Q,E_Q$), and a node $u\in V_Q$ with a potential match $v\in V$.
Let $BG$ be the bipartite graph that inspects all the \emph{LR} constraints defined over children (resp. parents) of $u$ 
w.r.t $v$. These \emph{LR} constraints are satisfied by some children (resp. parents) of $v$ iff there is a complete matching over $BG$. 
Moreover, this can be decided in at most $O(|V_Q||V|\sqrt{|V_Q|+|V|})$ time.
\end{my_theorem}

We emphasize that for each node $u$ in $Q$ and each potential match $v$ of $u$ in $G$, we construct at most two bipartite graphs, the first one to 
inspect \emph{LR} constraints that are defined over children of $u$, and the second one to inspects those defined over parents of $u$.

\begin{my_example}
As explained in Example \ref{example:LO_Properties_Satisfaction}, the \emph{LR} constraint defined over the children of $q_1$ in $Q$ is not 
satisfied by the children of its potential match $d_1$ in $G_1$. This is confirmed by the bipartite graph $BG_1$ of 
Fig. \ref{LO_Checking_Figure} (d) which has a maximum matching of size $2$ (does not cover the set $X$). 
Thus, no complete matching exists over $BG_1$ and, according to Theorem \ref{theorem:NS_Condition_For_LO}, we can conclude that the underlying 
\emph{LR} constraint is not satisfied by the children of $d_1$. Consider the bipartite graph $BG_2$ of Fig. \ref{LO_Checking_Figure} (e) 
that inspects the same \emph{LR} constraint w.r.t the potential match $d_1$ of $G_2$. Bold edges in $BG_2$ represent a maximum matching of 
size $3$. Thus, a complete matching exists over $BG_2$ which implies that the \emph{LR} constraint, defined over the children of $q_1$ in $Q$, 
is satisfied by the children of its potential match $d_1$ of $G_2$.
\end{my_example}

\section{An Algorithm for Triple Simulation}\label{section:Algorithm_for_TripleSimulation}
Our algorithm, referred to as \textsc{\textbf{TSim}}, is shown in the Fig. \ref{TSim_Algo}. Given a pattern graph $Q$ and a data 
graph $G$, \textsc{\textbf{TSim}}($Q, G$) returns the match result $\mathcal{M}_{T}(Q,G)$, if $Q\prec_T G$, and $\emptyset$ otherwise. 
This match result contains each subgraph of $G$ that satisfies all \emph{CPL} relationships of $Q$.

First, we compute for each node $u\in V_Q$, the set $\textsc{sim}(u)$ of all its potential matches 
in $V$ [lines 1-3]. In order to preserve efficiently the \emph{CPL} relationships of $Q$ over $G$, we define four auxiliary structures [line 4] as 
follows. For any node $u\in V_Q$, \textbf{CP}($Q,u$) contains all children and parents of $u$ that are concerned by \emph{Child} and/or 
\emph{Parent} relationships; and \textbf{LR}($Q,u$) contains those concerned by some \emph{LR} constraints. 
Moreover, for each potential match $v$ of $u$ in $G$, \textbf{\textsc{ChildAsMatch}}($Q,G,v,u$) returns the number of $v$'s children  
that are potential matches of $u$ in $G$ (i.e. each child $v^{'}$ of $v$ with $v^{'}\in\textsc{sim}(u)$); 
and \textbf{\textsc{ParentAsMatch}}($Q,G,v,u$) returns the number of $v$'s parents that are potential matches of $u$ in $G$.

Algorithm \textbf{TSim} preserves the \emph{Child} and \emph{Parent} relationships of $Q$ [lines 6-15] as follows. 
Given a node $u\in V_Q$, a potential match $v$ of $u$ is kept in $\textsc{sim}(u)$ unless: 1) $u$ has a child $u^{'}\in\textbf{CP}(Q,u)$ 
but $v$ has no child that matches $u^{'}$ (i.e. \textbf{\textsc{ChildAsMatch}}($Q,G,v,u^{'}$)=0); or 2) 
$u$ has a parent $u^{'}\in\textbf{CP}(Q,u)$ but $v$ has no parent that matches $u^{'}$ 
(i.e. \textbf{\textsc{ParentAsMatch}}($Q,G,v,u^{'}$)=0). If one of these two conditions is satisfied then $v$ is an incorrect match of $u$, 
w.r.t duality properties, and is removed from $\textsc{sim}(u)$ [lines 8 + 13].
The checking of \emph{LR} constraints [lines 17-19] is done through the procedure \textbf{LR\_Checking}. 
Given a node $u\in V_Q$ with a potential match $v\in V$. 
According to Definition \ref{definition:Bipartite_Graph_Inspecting_LO}, the procedure \textbf{LR\_Checking} constructs 
two bipartite graphs: $BG_1$ that inspects all the \emph{LR} constraints defined over the children of $u$ [lines 2-5]; and  
$BG_2$ that inspects those defined over the parents of $u$ [lines 6-9]. 
If a complete matching exists over $BG_1$ and another one exists over $BG_2$ then, according to Theorem \ref{theorem:NS_Condition_For_LO}, 
we conclude that: \textit{a)} all the \emph{LR} constraints defined over the children of $u$ are satisfied by some children of $v$; 
and \textit{b)} all the \emph{LR} constraints defined over the parents of $u$ are satisfied by some parents of $v$. 
Thus, the procedure returns $true$ only if these two complete matching exist over $BG_1$ and $BG_2$. 
If the procedure returns $false$ then there is at least one \emph{LR} constraint defined over the children (resp. parents) of $u$ which is not 
satisfied by the children (resp. parents) of $v$. In this case, $v$ is an incorrect match of $u$, w.r.t \emph{LR} constraints, and is removed 
from $\textsc{sim}(u)$ [line 18]. 
The procedure \textbf{\textit{CompleteMatch}}\footnote{This procedure finds the maximum matching over $BG_1$ (resp. $BG_2$), 
using the algorithm of Hopcroft et al. \cite{Hopcroft73}, and then checks whether the size of this maximum matching is equals to $|X_1|$ (resp. $|X_2|$).}
is an implementation of the algorithm of Hopcroft and Karp \cite{Hopcroft73}.

Each time a data node $v$ is removed from $\textsc{sim}(u)$, the cardinalities stored by the structures 
\textbf{\textsc{ChildAsMatch}} and \textbf{\textsc{ParentAsMatch}} are updated according to the couple $(u,v)$. This is done by the procedure 
\textbf{\textsc{UpdateStruct}}.
The two phases discussed above (checking of \emph{duality} properties and \emph{LR} constraints) are repeated by algorithm \textbf{TSim} until 
there are no more changes [lines 5-22]. Finally, the maximum match relation $S_{T}$ that corresponds to 
Definition \ref{Definition_Of_TripleSimulation} is defined, and its corresponding match result $\mathcal{M}_{T}(Q,G)$ is constructed and 
returned.

\begin{my_theorem}\label{theorem:TSim_Time_Complexity}
For any pattern graph $Q$($V_Q,E_Q$) and data graph $G$($V,E$), algorithm \textbf{TSim} takes at most
$O(|Q||G|+|V_Q|^{3}|V|^{2}\sqrt{|V_Q|+|V|})$ time to decide whether $Q\prec_T G$ and to find the 
match result $\mathcal{M}_{T}(Q,G)$. Moreover, it takes $O(|Q||G|)$ time if $Q$ has no \textit{LR} constraint.\footnote{Given a graph $G$($V,E$), $|G|=|V|+|E|$.}
\end{my_theorem}

The worst-case time complexity of \textbf{TSim} is bounded by $O(|Q|^{2}|G|^{1.5})$. 
As opposed to the \textsc{NP-Completeness} of its traditional counterpart via subgraph isomorphism \cite{Counting_Quantifiers_Fan_16}, triple 
simulation allows to match pattern graphs with \emph{LR} constraints in polynomial-time.

\begin{figure}[!htbp]
\rule{\linewidth}{0.5pt}
\small
\begin{flushleft}
\textbf{Algorithm} \textsc{\textbf{TSim}}($Q$, $G$)\newline
\textit{Input}: Graph pattern $Q$($V_Q,E_Q,\lambda_Q$), data graph $G$($V,E,\lambda$).\newline
\textit{Output}: The match result $\mathcal{M}_{T}(Q,G)$ if $Q\prec_T G$ and $\emptyset$ otherwise.
\end{flushleft}
\begin{algorithmic}[1]
\ForAll{$u \in V_Q$}\hfill\textbf{/*} \textsc{Potential matches of each node in $Q$} \textbf{*/}
    \State \textsc{sim($u$)} := \{$v$ $|$ $v\in V$ and $\lambda_Q(u)$=$\lambda(v)$\};
\EndFor

\State \textbf{initAuxStruct($Q,G$)};

\Do
\ForAll{$(u,v)$ \textbf{with} $v\in$ \textsc{sim($u$)}}
    \ForAll{\textbf{child} $u^{'}$ of $u$ \textbf{with} $u^{'}\in\textbf{\textsc{CP}}(Q,u)$}\hfill\textbf{/*} \textsc{Preserving \emph{Child} relations} \textbf{*/}
	\If{(\textbf{\textsc{ChildAsMatch($Q,G,v,u^{'}$)}}$=0$)}
	    \State \textsc{sim($u$)} := \textsc{sim($u$)}$\setminus\{v\}$; \textbf{UpdateStruct($G,u,v$)};
	\EndIf
    \EndFor
    \ForAll{\textbf{parent} $u^{'}$ of $u$ \textbf{with} $u^{'}\in\textbf{\textsc{CP}}(Q,u)$}\hfill\textbf{/*} \textsc{Preserving \emph{Parent} relations} \textbf{*/}
	\If{(\textbf{\textsc{ParentAsMatch($Q,G,v,u^{'}$)}}$=0$)}
	    \State \textsc{sim($u$)} := \textsc{sim($u$)}$\setminus\{v\}$; \textbf{UpdateStruct($G,u,v$)};
	\EndIf
    \EndFor
    \If{(\textbf{\textsc{LR\_Checking}}($Q, G, u, v$)=$false$)}\hfill\textbf{/*} \textsc{Preserving \emph{LR} constraints} \textbf{*/}
	\State \textsc{sim($u$)} := \textsc{sim($u$)}$\setminus\{v\}$; \textbf{UpdateStruct($G,u,v$)};
    \EndIf
    \IIf{(\textsc{sim}($u$) = $\emptyset$)}
	\Return $<\emptyset,\emptyset>$ ;
    \EndIIf
\EndFor

\doWhile{there are changes;}

\State $S_{T}$ := \{($u$, $v$) $|$ $u\in V_Q$ and $v\in$ \textsc{sim}($u$)\};

\State \textit{Construct the match result} $\mathcal{M}_{T}(Q,G)$ that corresponds to $S_{T}$;

\State \Return $\mathcal{M}_{T}(Q,G)$;
\normalsize
\end{algorithmic}

\small
\begin{flushleft}
\textbf{Procedure} \textsc{\textbf{UpdateStruct}}($Q,G,u,v$)\newline
\textit{Input}: A pattern graph $Q$, data graph $G$($V,E$), a query node $u$ with a removed potential match $v$.\newline
\textit{Output}: Updates the auxiliary structures \textbf{ChildAsMatch} and \textbf{ParentAsMatch}.
\end{flushleft}
\begin{algorithmic}[1]
\State \textbf{Do} \textbf{\textsc{ChildAsMatch}}($Q,G,v^{'},u$) := \textbf{\textsc{ChildAsMatch}}($Q,G,v^{'},u$) - 1 \textbf{for each} $(v^{'},v)\in E$;
\State \textbf{Do} \textbf{\textsc{ParentAsMatch}}($Q,G,v^{'},u$) := \textbf{\textsc{ParentAsMatch}}($Q,G,v^{'},u$) - 1 \textbf{for each} $(v,v^{'})\in E$;
\normalsize
\end{algorithmic}

\small
\begin{flushleft}
\textbf{Procedure} \textsc{\textbf{\textsc{LR\_Checking}}}($Q, G, u, v$)\newline
\textit{Input}: Graph pattern $Q(V_Q,E_Q)$, data graph $G(V,E)$, a node $u\in V_Q$ with a potential match $v\in V$.\newline
\textit{Output}: Whether all the \emph{LR} constraints defined over $u$ are satisfied by children and/or parents of $v$.
\end{flushleft}
\begin{algorithmic}[1]
  \State \textit{\textbf{BG$_1$}} := $(X_1 \cup Y_1,E_1)$; \textit{\textbf{BG$_2$}} := $(X_2 \cup Y_2,E_2)$; where $X_1=Y_1=X_2=Y_2=E_1=E_2=\emptyset$;
  \ForAll{\textbf{child} $u^{'}$ of $u$ \textbf{with} $u^{'}\in \textbf{\textsc{LR}}(Q,u)$}
      \State $X_1$ := $X_1$ $\cup$ \{$u^{'}$\};
      \State \textbf{Do} $Y_1$ := $Y_1$ $\cup$ \{$v^{'}$\}; $E_1$ := $E_1$ $\cup$ \{($u^{'},v^{'}$)\}; \textbf{for each} ($v^{'}\in$ \textsc{sim}($u^{'}$) \textbf{with} $(v,v^{'})\in E$); 
  \EndFor
  
  \ForAll{\textbf{parent} $u^{'}$ of $u$ \textbf{with} $u^{'}\in \textbf{\textsc{LR}}(Q,u)$}
      \State $X_2$ := $X_2$ $\cup$ \{$u^{'}$\};
      \State \textbf{Do} $Y_2$ := $Y_2$ $\cup$ \{$v^{'}$\}; $E_2$ := $E_2$ $\cup$ \{($u^{'},v^{'}$)\}; \textbf{for each} ($v^{'}\in$ \textsc{sim}($u^{'}$) \textbf{with} $(v^{'},v)\in E$);
  \EndFor
  
  \State \Return $true$ \textbf{if} (\textit{\textbf{CompleteMatch}}(\textit{\textbf{BG$_1$}}) \textbf{\&} \textit{\textbf{CompleteMatch}}(\textit{\textbf{BG$_2$}})); 
  \textbf{and $false$ otherwise};
  
\normalsize
\end{algorithmic}

\rule{\linewidth}{0.5pt}
\caption{Algorithm for Triple Simulation.}
\label{TSim_Algo}
\end{figure}

\section{Triple Simulation with Locality}\label{section:TSim_With_Locality}
The next example suggests to incorporate the notion of locality \cite{Fan14} into our algorithm \textbf{TSim} in order to overcome excessive 
matching and thus to improve the quality of our match results.

\begin{my_example}\label{example:TSim_With_Locality}
Consider the graphs depicted in Fig. \ref{Introductory_Example_Figure}. We 
extend the subgraph $G_1$ with the following relationships: $d_1\leftarrow d_{13}\leftarrow d_7$ where $d_{13}$ is a new node labeled 
with \textbf{SE}. Let $G_{1}^{'}$ be the subgraph that results from this modification.
When triple simulation is adopted, \textbf{TSim} returns $G_{1}^{'}$ as the only match of $Q_1$ in $G$. 
The \textbf{BIO} found in $G_{1}^{'}$ (node $d_2$) is recommended by two \textbf{SE} ($d_8$ and $d_{13}$) as specified by $Q_1$. 
However, \textbf{TSim} returns an excessive match of the cycle $\textbf{AI}\leftrightarrows\textbf{DM}$, 
i.e. the cycle $d_9\rightarrow\dots\rightarrow d_{12}\rightarrow d_9$ in $G_{1}^{'}$, that one does not want.
\end{my_example}

Next is a new definition of triple simulation that takes into account the notion of locality.

\begin{my_definition}\label{definition:Triple_Simulation_With_Locality}
A data graph $G$ matches a pattern graph $Q$ via \emph{triple simulation} and \emph{under locality}, denoted $Q\prec^{L}_{T} G$, if there exists a 
subgraph $G_s$ of $G$ centered at some node $v$ s.t.: 
\begin{enumerate}
 \item the radius of $G_s$ is bounded by $d_Q$, i.e., for each node $v^{'}$ in $G_s$, \emph{dist}($v,v^{'}$)$\leq d_Q$; and 
 \item $Q\prec_T G_s$ with the maximum match relation $S_T$.
\end{enumerate}
The match result $\mathcal{M}^{L}_{T}(Q,G)$ is defined with $\bigcup_{i}\mathcal{M}_{T}(Q,G_i)$ where each $G_i$ is a subgraph of $G$ that 
satisfies the previous conditions.
\end{my_definition}

To implement the Definition \ref{definition:Triple_Simulation_With_Locality}, one can replace only the procedure \textbf{dualSim} in the 
algorithm \textbf{Match} \cite{Fan14} with our algorithm \textbf{TSim}. 
Let \textbf{Match$^{+}$} be the algorithm that results from this combination. 
Given a data graph $G$ and a pattern graph $Q$. Algorithm \textbf{Match$^{+}$}\footnote{Not given here since its definition is trivial.} 
extracts a subgraph $G_v$ over each node $v$ in $G$, provided that its radius does not exceed $d_Q$. It then matches $G_v$ over $Q$ via triple 
simulation (instead of dual simulation). The match found on each subgraph has a reasonable size and 
satisfies all the \emph{CPL} relationships of $Q$.

\begin{my_theorem}\label{theorem:MatchPlus_Time_Complexity}\footnote{This result is a combination of Theorem \ref{theorem:TSim_Time_Complexity} and Theorem 4.1 of 
Ma et al. \cite{Fan14}.}
For any pattern graph $Q$($V_Q,E_Q$) and data graph $G$($V,E$), algorithm \textbf{Match$^{+}$} takes 
at most $O(|V|^{2}+|Q||G||V|+|V_Q|^{3}|V|^{3}\sqrt{|V_Q|+|V|})$ time to decide whether $Q\prec_T^{L} G$ and to find the corresponding match 
result $\mathcal{M}^{L}_{T}(Q,G)$.
\end{my_theorem}

The complexity of \textbf{Match$^{+}$} is bounded by $O(|Q|^{2}|G|^{2})$ while that of \textbf{Match}\cite{Fan14} is bounded by 
$O(|Q||G|^{2})$. This promises that combining our results with existing orthogonal approaches will not increase drastically the complexity of 
graph pattern matching.

\section{Conclusion}
We have discussed pattern graphs with \emph{LR} constraints that existing approaches do not preserve 
\cite{Fan14, GPM_From_Intractable_To_Polynomial_Time} or preserve in exponential time \cite{Counting_Quantifiers_Fan_16}. 
To tackle this NP-Completeness, we have showed that \emph{LR} constraints can be preserved in polynomial-time when treated as maximum 
matching in bipartite graphs, and we proposed an algorithm to implement this result. 

We are to stduy other constraints that can be preserved in polynomial-time, e.g., \emph{negation} and \emph{optional edges}.
The polynomial-time of our algorithm may make graph pattern matching infeasible when conducted on graphs with millions of nodes and billions of 
edges (e.g. Facebook \cite{FacebookStatistic}). To boost the matching on large data graphs, we plan to extend our work with some optimization 
techniques: \textit{1) incremental graph pattern matching} \cite{Incremental_GPM}, 
\textit{2) pattern matching on distributed data graphs} \cite{GPM_On_Ditributed_Graphs_1,GPM_On_Ditributed_Graphs_2,GPM_On_Ditributed_Graphs_3}, 
and \textit{3) pattern matching on compressed data graphs} \cite{Preserving_Graph_Compression_1, Preserving_Graph_Compression_2}. These 
techniques are orthogonal, but complementary, to our work.

\bibliography{TSim}

\newpage
\noindent\textbf{APPENDIX}
\appendix
\section{Proof of Theorem \ref{theorem:NS_Condition_For_LO}.}

\noindent\textbf{Theorem 1. (Recall)}
Given a data graph $G$($V,E$), a pattern graph $Q$($V_Q,E_Q$), and a node $u\in V_Q$ with a potential match $v\in V$.
Let $BG$ be the bipartite graph that inspects all the \emph{LR} constraints defined over children (resp. parents) of $u$ 
w.r.t $v$. These \emph{LR} constraints are satisfied by some children (resp. parents) of $v$ iff there is a complete matching over $BG$. 
Moreover, this can be decided in at most $O(|V_Q||V|\sqrt{|V_Q|+|V|})$ time.\newline

To simplify the proof, we consider only the case of \emph{LR} constraints defined over children of $u$. The second case, i.e. 
when parents of $u$ are concerned by some \emph{LR} constraints, can be studied in the same way.
Satisfying \emph{LR} constraints is closer to the problem of \emph{perfect matching in bipartite graph} \cite{Cormen}, or moreover,
a \emph{System of Distinct Representatives} \cite{OnRepresentativeSubsets}. In our case, node sets $X$ and $Y$ of our bipartite graphs have not the 
same size then we use the term of \emph{complete matching} instead of \emph{perfect matching}. 
Given a bipartite graph $B$=($X\cup Y, Z$). A maximum matching 
$S\subseteq Z$ is the largest subset of the edge set $Z$ such that no two edges start/end at the same node. If $S$ is a \emph{complete matching}, 
i.e. $|S|=|X|$, then for each node $x\in X$ there is one and only one edge $(x,y)\in S$ that connects it with a node $y\in Y$. 
We say that all elements of $X$ are covered (i.e. matched).

\noindent\textit{\textbf{\boldmath{$\Longrightarrow$}}} Consider that all \emph{LR} constraints defined over children of $u$ are 
satisfied by some children of its potential match $v$. 
Recall that $BG$ is defined with $(X\cup Y, Z)$ where $X$ contains each child of $u$ that is concerned by an \emph{LR} constraint; and $Y$ 
contains each child of $v$ that matches at least one child of $u$ in $X$. Let $K$ be the number of $u$'s children that are concerned by \emph{LR} 
constraints (i.e. $K=|X|$). Since all the \emph{LR} constraints in question are satisfied by some children of $v$ then, 
for each single one defined over the subset 
$C_u=u_1,\dots,u_N$ of $N$ children of $u$ ($2\leq N\leq K$), $v$ satisfies condition (2) of Definition \ref{definition:LR_Constraints} and has a 
subset $C_v=v_1,\dots,v_N$ of children such that each $v_i$ matches a child $u_i$ of $u$. 
Notice that two different \emph{LR} constraints are defined with two different labels, thus children of $v$ that satisfy one \emph{LR} constraint 
are different from those that satisfy another \emph{LR} constraint. 
By following the same principle, to satisfy all \emph{LR} constraints defined over children of $u$, $v$ has certainly $K$ distinct children such 
that each one is matched to only one child of $u$ which is concerned by some \emph{LR} constraint. 
This matching can be represented by $K$ edges that connect each child of $v$ in $Y$ to only one child of $u$ in $X$ (\textbf{*}). Moreover, if 
two children of $v$ has the same label then they are concerned by the same \emph{LR} constraint and, according to Definition \ref{definition:LR_Constraints}, 
are matched to different nodes in $X$ (\textbf{**}). From (\textbf{*}) and (\textbf{**}), 
we conclude that these $K$ edges do not start/end at the same node and then represent a complete matching over the bipartite graph $BG$. 
Therefore, if all \emph{LR} constraints defined over children of $u$ are satisfied by some 
children of $v$, then there is a complete matching over the bipartite graph $BG$ that inspects these \emph{LR} constraints w.r.t $v$.

\noindent\textit{\textbf{\boldmath{$\Longleftarrow$}}} Consider that there is a complete matching over the bipartite graph $BG$. 
According to our definition of complete matching, there is an edge that connects each node in $X$ (i.e. a child $u^{'}$ of $u$ that is concerned 
by an \emph{LR} constraint) to only one node in $Y$ (i.e. a child $v^{'}$ of $v$ with $v^{'}\in\textsc{sim}(u^{'})$), and moreover, 
each node in $Y$ is connected to only one node in $X$. We conclude that $v$ has at least $K$ children ($K=|X|$) and there exists an order over these 
children that allows to match each one to only one child of $u$ which is concerned by some \emph{LR} constraint. Therefore, 
according to Definition \ref{definition:LR_Constraints}, each \emph{LR} constraint defined over some children of $u$ is satisfied by some children 
of $v$.

The node set $X$ (resp. $Y$) of the bipartite graph $BG$ may have at most $|V_Q|$ (resp. $|V|$) nodes. Moreover, the edge set $Z$ may have at most 
$|V_Q||V|$ edges. To check whether there exists a complete matching over $BG$, we look first for the maximum matching over $BG$ and we then check 
whether its cardinality is equals to $|X|$. The best algorithm to find a maximum matching over a bipartite graph with node set $N$ and edge set $M$, 
discovered by Hopcroft and Karp \cite{Hopcroft73}, runs in $O(|M|\sqrt{|N|})$ time. 
Thus, by using this algorithm, the necessary and sufficient condition of Theorem \ref{theorem:NS_Condition_For_LO} can be checked in at most 
$O(|V_Q||V|\sqrt{|V_Q|+|V|})$ time.

\section{Proof of Theorem \ref{theorem:TSim_Time_Complexity}.}

\noindent\textbf{Theorem 2. (Recall)}
For any pattern graph $Q$($V_Q,E_Q$) and data graph $G$($V,E$), algorithm \textbf{TSim} takes at most
$O(|Q||G|+|V_Q|^{3}|V|^{2}\sqrt{|V_Q|+|V|})$ time to decide whether $Q\prec_T G$ and to find the 
match result $\mathcal{M}_{T}(Q,G)$. Moreover, it takes $(|Q||G|)$ time in the absence of \emph{LR} constraints.\newline

Given a pattern graph $Q$($V_Q,E_Q,\lambda_Q$) and a data graph $Q$($V,E,\lambda$). 
It takes $O(|V_Q||V|)$ time to compute \textsc{sim} sets for all query nodes of $Q$ [lines 1-3]. We define each \textsc{sim}($u$) as an indexed 
structure which allows, in constant time, 1) to check whether some data node $v$ belongs to $\textsc{sim}(u)$; or 2) to remove it from 
$\textsc{sim}(u)$.\newline

\noindent\textit{(A}) The auxiliary structures \textbf{CP} and \textbf{LR} can be constructed in at most $O(|V_Q|^{2})$ time as follows. 
For any node $u\in V_Q$, we define an indexed list \textbf{LabelOcc}($u,l$) which returns the number of children of 
$u$ that are labeled with $l$. This list can be constructed in $O(|V_Q|)$ time by parsing all children of $u$.
For each child $u^{'}$ of $u$, 
if \textbf{LabelOcc}($u,\lambda_Q(u^{'})$)$>1$, then other children of $u$ have the same label as $u^{'}$. Thus, $u^{'}$ is 
concerned by an \emph{LR} constraint and must belong to \textbf{LR}($Q,u$). Otherwise, i.e. \textbf{LabelOcc}($u,\lambda_Q(u^{'})$)$=1$, 
$u^{'}$ is the unique child of $u$ that has the label $\lambda_Q(u^{'})$ and thus must belong to \textbf{CP}($Q,u$). This process is repeated 
similarly over parents of $u$ to complete the definition of \textbf{CP}($Q,u$) and \textbf{LR}($Q,u$). It is clear that for each node $u\in V_Q$, 
\textbf{CP}($Q,u$) and \textbf{LR}($Q,u$) can be constructed in $O(|V_Q|)$ time. Therefore, for all nodes of $Q$, the cost becomes $O(|V_Q|^{2})$.

\noindent\textit{(B}) It is easy to verify that for each query node $u\in V_Q$ and data node $v\in V$, 
\textbf{ChildAsMatch}($Q,G,v,u$) (resp. \textbf{ParentAsMatch}($Q,G,v,u$)) can be constructed in $O(|V|)$ time by parsing each child 
(resp. parent) of $v$ and checking, in constant time, if this child belongs to $\textsc{sim}(u)$. 
Therefore, by considering all nodes of $Q$ and $G$, the structures \textbf{ChildAsMatch} and \textbf{ParentAsMatch} can 
be constructed in at most $O(|V_Q||V|^{2})$ time.

\noindent\textit{(C}) In addition to the four auxiliary structures described above, we construct in $O(|E|)$ time (resp. $O(|E_Q|)$ time) an indexed structure over the 
edges of $E$ (resp. $E_Q$) in order to check in constant time whether some data edge (resp. query edge) exists. 
Moreover, we define sets of children and parents of each query node $u\in V_Q$ (resp. data node $v\in V$) which can be done in $O(|E_Q|)$ time 
(resp. $O(|E|)$ time).

From \textit{(A)}, \textit{(B)} and \textit{(C)}, we conclude that the cost of the call \textbf{initAuxStruct}($Q,G$) [line 4] 
remains bounded by $O(|V_Q||V|^{2})$.\newline

Each time we remove some data node $v$ from $\textsc{sim}(u)$, the procedure \textbf{UpdateStruct}($u,v$) of Fig. \ref{TSim_Algo} takes 
$O(|V|)$ time 
to update the structures \textbf{ChildAsMatch} and \textbf{ParentAsMatch}. This remove operation can be done at most $|V_Q||V|$ time. Thus, 
the lines [9+14+18] of algorithm \textbf{TSim} take at most $O(|V_Q||V|^{2})$ time.\newline

Given a query node $u$ with a potential match $v$. The checking of \emph{Child} relationships [lines 7-11], as well as \emph{Parent} 
relationships [lines 12-16] is done in at most $O(|V_Q|)$ time by using the indexed structures \textbf{ChildAsMatch} and \textbf{ParentAsMatch} 
(inspired from \cite{GraphSimulation}). Recall that the cost necessary to update these indexed structures is computed separately.

\begin{figure}
\rule{\linewidth}{0.5pt}
\small
\begin{flushleft}
\textbf{Procedure} \textsc{\textbf{MatchResult}}($Q$, $G$, $S_{T}$)\newline
\textit{Input}: A pattern graph $Q(V_Q,E_Q,\lambda_Q)$, a data graph $G(V,E,\lambda)$, and the maximum match relation $S_{T}$ for which $Q\prec_T G$.\newline
\textit{Output}: The match result $\mathcal{M}_{T}(Q,G)$ that corresponds to $S_{T}$.
\end{flushleft}
\begin{algorithmic}[1]
\State $\mathcal{M}_{T}(Q,G)$ := ($V_r,E_r,\lambda_r$);\hfill\textbf{/*} \textsc{A disconnected graph} \textbf{*/}

\ForAll{$(u,v)\in S_{T}$}
    \State $V_r$ := $V_r\cup \{v\}$; $\lambda_r(v)$ := $\lambda(v)$;
\EndFor

\ForAll{\textbf{edge} $(u,u^{'})\in E_Q$}
    \ForAll{$(u,v)\in S_{T}$ \textbf{and} $(u^{'},v^{'})\in S_{T}$}
        \IIf{$(v,v^{'})\in E$}
            $E_r$ := $E_r\cup \{(v,v^{'})\}$;
        \EndIIf
    \EndFor
\EndFor

\State \Return $\mathcal{M}_{T}(Q,G)$;
\normalsize
\end{algorithmic}
\rule{\linewidth}{0.5pt}
\caption{Procedure to construct Match Results.}
\label{procedure:MatchResult}
\end{figure}

The call \textbf{LR\_Checking}($Q,G,u,v$) [line 17] is done in at most $O(|V_Q||V|\sqrt{|V_Q|+|V|})$ time as we explain 
hereafter. As depicted by the procedure \textbf{LR\_Checking} of Fig. \ref{TSim_Algo}, 
we construct first two bipartite graphs $BG_1$ and $BG_2$ that 
inspect the \emph{LR} constraints defined over children of $u$ [lines 2-5] and those 
defined over parents of $u$ respectively [lines 6-9]. 
We get all children/parents of $u$ in at most $O(|V_Q|)$ time by using our precomputed sets of children and parents. 
Thus, the construction of $BG_1$ as well as $BG_2$ requires a time bounded by $O(|V_Q||V|)$. Next, we use the procedure 
\textbf{\textit{CompleteMatch}} (not detailed here) to check whether there exist two complete matchings over $BG_1$ and $BG_2$ respectively. 
Our bipartite graphs have at most $|V_Q\cup V|$ nodes and $|V_Q||V|$ edges. According to Theorem \ref{theorem:NS_Condition_For_LO}, 
the existence of complete matching over $BG_1$ and $BG_2$ can be checked in at most $O(|V||V_Q|\sqrt{|V_Q|+|V|})$ time. Therefore, 
the checking of \emph{LR} constraints by algorithm \textbf{TSim} [lines 17-19] requires a time bounded by $O(|V||V_Q|\sqrt{|V_Q|+|V|})$.

For a query node $u$ with a potential match $v$, the checking of duality properties takes $O(|V_Q|)$ time while that of 
\emph{LR} constraints takes $O(|V_Q||V|\sqrt{|V_Q|+|V|})$ time. This tells us that the worst case arises when children (resp. parents) of 
$u$ are concerned by only \emph{LR} constraints. The checking process [lines 6-21] over all potential matches of $u$ 
is done in at most $O(|V_Q||V|)$ time, in case of duality properties only, and in $O(|V_Q||V|^{2}\sqrt{|V_Q|+|V|})$ time in 
case of \emph{LR} constraints only.

Inspired from \cite{GraphSimulation}, the checking process (of duality properties and \emph{LR} constraints) [lines 5-22] is executed over the 
nodes of $Q$ in a \emph{deterministic} manner: first over a randomly-chosen query node $u$, 
after over adjacent nodes of $u$ (children and parents) and so on. 
In this way, each time some $\textsc{sim}$ set is changed we repeat the checking process over all already visited nodes since this change may 
influence on their $\textsc{sim}$ sets. Thus, the \textsc{Do-While} loop will repeat the checking process $|V_Q|$ 
times over each query node in $Q$.

The definition of the maximum match relation $S_{T}$ [line 23] can be done in at most $O(|V_Q||V|)$ time. 
The match result $\mathcal{M}_{T}(Q,G)$ that corresponds to $S_T$ can be defined in at most $O(|E_Q||E|)$ time [line 24]. 
To proof this cost, we give in Fig. \ref{procedure:MatchResult} the procedure \textbf{MatchResult} which defines the match result that corresponds 
to some maximum match relation. The first \textsc{For-Each} loop of this procedure takes $O(|V_Q||V|)$ time since the size of $S_T$ is 
bounded by $|V_Q||V|$. 
The second \textsc{For-Each} loop is repeated $|E_Q|$ time, and in each iteration, 
we make all combinations between children of $u$ and those of $u^{'}$, which 
takes $O(|V|^{2})$ time. We suppose that it can be checked in constant time whether $(u,v)\in S_T$ (resp. $(u^{'},v^{'})\in S_T$). Thus, 
the overall time complexity of the procedure \textbf{MatchResult} remains bounded by $O(|E_Q||V|^{2})$.\newline

Hereafter a summary of all the above-mentioned costs of algorithm \textbf{TSim}:

\begin{itemize}
 \item $O(|V_Q||V|)$ time to compute all $\textsc{sim}$ sets.
 \item $O(|V_Q||V|^{2})$ time for the call of \textbf{initAuxStruct}.
 \item $O(|V_Q||V|^{2})$ time for the calls of \textbf{UpdateStruct}.
 \item $O(|V_Q|^{3}|V|^{2}\sqrt{|V_Q|+|V|})$ time for checking of \emph{LR} constraints, and $O(|V_Q|^{3}|V|)$ time 
 for checking of duality properties.
 \item $O(|E_Q||V|^{2})$ time for the definition of the match result $\mathcal{M}_{T}(Q,G)$.
\end{itemize}

For any pattern graph $Q$($V_Q,E_Q$) and data graph $G$($V,E$) in practice, we have $|V_Q|<|V|$. Thus, the cost $|V_Q|^{3}|V|$ can be bounded by 
$|V_Q|^{2}|V|^{2}$. Moreover, $|E_Q|$ (resp. $|E|$) is bounded by $|V_Q|^{2}$ (resp. $|V|^{2}$). 

Finally, we conclude that the overall time complexity of algorithm \textbf{TSim} is bounded by $O(|Q||G|)$ is case of only duality properties, 
and by $O(|Q||G|+|V_Q|^{3}|V|^{2}\sqrt{|V_Q|+|V|})$ in presence of \emph{LR} constraints.

\section{Proof of Theorem \ref{theorem:MatchPlus_Time_Complexity}}

\noindent\textbf{Theorem 3. (Recall)}
For any pattern graph $Q$($V_Q,E_Q$) and data graph $G$($V,E$), algorithm \textbf{Match$^{+}$} takes 
at most $O(|V|^{2}+|Q||G||V|+|V_Q|^{3}|V|^{3}\sqrt{|V_Q|+|V|})$ time to decide whether $Q\prec_T^{L} G$ and to find the corresponding match 
result $\mathcal{M}^{L}_{T}(Q,G)$.\newline

Given a pattern graph $Q$($V_Q,E_Q$) and a data graph $G$($V,E$). Ma et al. \cite{Fan14} show that their algorithm \textbf{Match} requires 
$O(|V|(|V|+ \textbf{\textit{DualSimCost}}))$ time to decide whether $Q\prec_{D}^{L} G$ and to find the match result $\mathcal{M}^{L}_{D}(Q,G)$. 
Here \textbf{\textit{DualSimCost}} denotes the cost of dual simulation since they preserve only child and parents relationships besides the notion 
of locality. Recall that \textbf{Match$^{+}$} is a new version of \textbf{Match} that we propose in order to take advantage of triple simulation as 
well as of locality, and this by replacing the dual simulation in algorithm \textbf{Match} with triple simulation. More precisely, we replace 
the procedure \textbf{DualSim} in algorithm \textbf{Match} with our algorithm \textbf{TSim}. 
Therefore, to get the result of Theorem \ref{theorem:MatchPlus_Time_Complexity}, one can replace intuitively \textbf{\textit{DualSimCost}} 
with the overall cost of algorithm \textbf{TSim} (Theorem \ref{theorem:TSim_Time_Complexity}).

\section{Discussion about Quantified Graph Patterns}\label{appendix_TSim_CQs}
Authors of \cite{Counting_Quantifiers_Fan_16} propose a new extension of subgraph isomorphism by supporting simple counting 
quantifiers (\emph{CQs}) on edges. These \emph{CQs} can express universal and existential quantification, numeric and ratio aggregate, as well as 
negation.

\begin{example}
The pattern graph \scriptsize{$Pr\xrightarrow{=100\%} PhD\rightarrow Conf\_Paper$}\normalsize$~$ looks for each professor 
such that all her PhD students (\emph{universal quantification}) have at least one conference paper (\emph{existential quantification}). 
The pattern graph \scriptsize{$Pr\xrightarrow{\geq 60\%} PhD\xrightarrow{\geq 2} Conf\_Paper\xrightarrow{=0} DBLP$}\normalsize $~$ looks for each 
professor such that $60\%$ of her PhD students (\textit{aggregate ratio}) have at least two conference papers (\textit{numeric ratio}) 
that are not indexed in DBLP (\emph{negation}).
\end{example}

\begin{definition}
A pattern graph with counting quantifiers, called \emph{quantified pattern graph} (\textbf{QGP}), 
is defined with $Q$($V,E,\lambda,\mathcal{C}$) where $V$, $E$, and 
$\lambda$ are the same as their conventional counterparts; and $\mathcal{C}$ is a function such that, for each edge $e\in E$, $\mathcal{C}(e)$ is 
given by: ``$=0$'', ``$=100\%$'', ``$\geq p \%$'', or ``$\geq p$'' ($p\geq 1$).
\end{definition}

Remark that conventional pattern graphs are a special case where for each edge $e$, $\mathcal{C}(e)\geq 1$ (only existential quantification). 
We omit $\mathcal{C}(e)$ from each edge $e$ if it is an existential quantification.

It is clear to see that \emph{LR} constraints are much close to \emph{counting quantifiers with numeric aggregate} 
(denoted shortly \textit{CQs$^{+}$}). Hence, we conduct in the next a comparison between \emph{LR} constraints and \textit{CQs$^{+}$} and we 
show how to extend our algorithm \textbf{\textsc{TSim}} to handle pattern graphs with \textit{CQs$^{+}$}. Since the other 
forms of \emph{CQs} are not too close to our problem, we consider in the next quantified pattern graphs with only numeric aggregates.

\begin{figure}[t]
\centering
   \includegraphics[width=300px,height=200px]{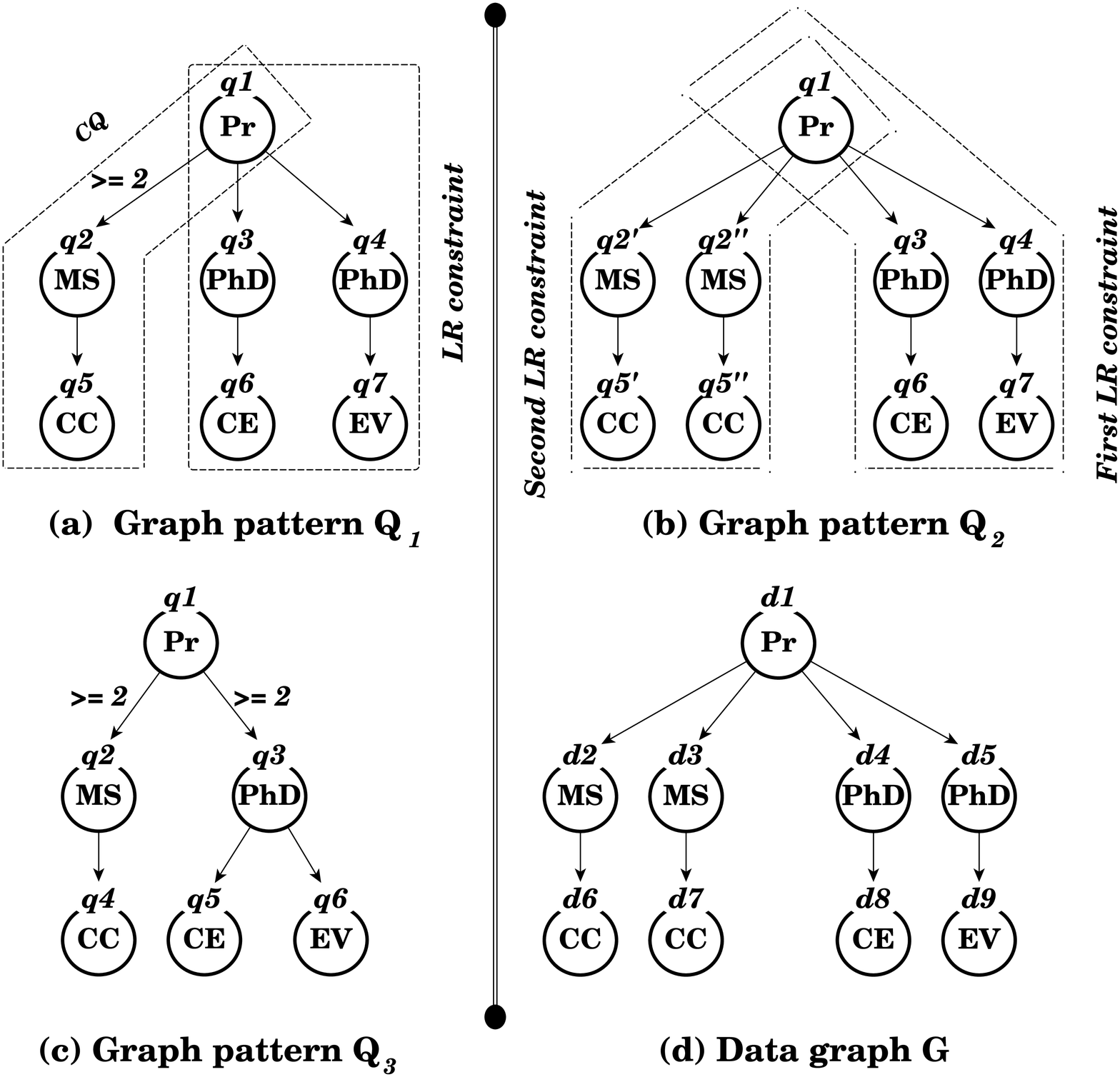}
   \caption{\emph{LR} constraints versus \textit{CQs$^{+}$}.}
   \label{LO_VS_CQ}
\end{figure}

\subsection{\emph{LR} Constraints v.s. CQs$^{+}$}
The limit of \textit{CQs$^{+}$} is that they specify the minimum number of children which must have all the same properties (\textit{child clone}). 
Formally, given the edge $A\xrightarrow{\geq p}B$ of some pattern graph $Q$. This specifies that: 1) each data node $v$, that matches $A$, 
must have \emph{at least} $p$ child nodes that match $B$; and 2) \emph{all} these $p$ nodes must satisfy \emph{the same properties set} that 
are defined over $B$ in $Q$. Moreover, \textit{CQs$^{+}$} are defined over children only.

An \emph{LR} constraint, however, specifies the minimum number of children or parents that has some query node such that they have all the same 
label but not necessarily the same properties. 
In addition, any \textit{CQ$^{+}$} can be transformed to an \emph{LR} constraint (Proposition \ref{proposition:Translate_CQs_Into_LO_Prop}), but 
the inverse is not always possible as shows the next example.

\begin{example}
Consider the pattern graphs $Q_1$, $Q_2$, $Q_3$, and the data graph $G$ depicted in Fig. \ref{LO_VS_CQ}. 
The pattern graph $Q_1$ looks for each professor (\textbf{Pr}) which has supervised: 1) at least two 
master students (\textbf{MS}) who have worked in the \emph{Cloud Computing} (\textbf{CC}) area; and 2) at least two PhD students who had topics 
related to \emph{Collaborative Editing} (\textbf{CE}) and \emph{Electronic Vote} (\textbf{EV}) respectively. 
Remark that $Q_1$ is composed by one \emph{LR} property and one \textit{CQ$^{+}$}. This \textit{CQ$^{+}$} can be easily replaced by an \emph{LR} 
constraint as follows: we replace the global child node $q_2$ by two copies of it, $q_2^{'}$ and $q_2^{''}$, such that the properties defined in 
$Q_1$ over $q_2$ (child $q_5$ of type \textbf{CC}) must be duplicated over each copy of it. This transformation yields for the pattern graph $Q_2$. 
See that $Q_1$ and $Q_2$ are equivalent: by using the algorithm in \cite{Counting_Quantifiers_Fan_16}, the matching of $Q_1$ over 
$G$ returns the whole data graph $G$ as match result, which is the same result returned by \textbf{\textsc{TSim}}($Q_2,G$).
However, it is clear that the \emph{LR} constraint of $Q_1$ can not be replaced by the \textit{CQ$^{+}$}  
``\scriptsize{$Pr\xrightarrow{\geq 2}PhD$}\normalsize'' as done with the pattern graph $Q_3$. Thus, $Q_1$ and $Q_3$ are not equivalent: 
matching $Q_3$ over $G$ with the algorithm in \cite{Counting_Quantifiers_Fan_16} yields  for an empty set.
\end{example}


\begin{proposition}\label{proposition:Translate_CQs_Into_LO_Prop}
Given a quantified pattern graph $Q$($V_Q,E_Q,\lambda_Q,\mathcal{C}$), a data graph $G$($V,E$), and a node $u\in V_Q$ with a potential match 
$v\in V$. Each \textit{CQ$^{+}$} defined with $\mathcal{C}$ over some child of $u$ can be transformed into an \emph{LR} constraint. 
Moreover, children of $v$ satisfy this \textit{CQ$^{+}$} iff they satisfy its equivalent \emph{LR} constraint.
\end{proposition}

In the following, we give another definition of triple simulation that takes into account \emph{CPL} relationships as well as \emph{CQs$^{+}$}. 
We show later that implementing this new definition requires just a simple extension of algorithm \textbf{TSim}.

\begin{my_definition}\label{Definition_Of_TripleSimulation}
Given a data graph $G(V,E,\lambda)$ and a quantified pattern graph $Q(V_Q,E_Q,\lambda_Q,\mathcal{C})$ where $\mathcal{C}$ defines only 
\textit{CQs$^{+}$}. Then, $G$ matches $Q$ via \emph{triple simulation}, denoted by $Q \prec_T G$, if there exists a binary match relation 
$S_{T}\subseteq V_Q \times V$ s.t.:

\begin{enumerate}
    \item For each $(u,v)\in S_{T}$, $\lambda_Q(u)=\lambda(v)$.
    \item For each $u\in V_Q$ there exists $(u,v)\in S_{T}$.
    \item For each $(u,v)\in S_{T}$ and for all simple edges $(u, u_{1}),...,(u, u_{n})\in E_Q$, there exists 
    \emph{\textbf{at least $n$ distinct children}} $v_{1},...,v_{n}$ of $v$ in $G$ such that: $(u_{1}, v_{1}),...,(u_{n}, v_{n})\in S_{T}$.
    \item For each $(u,v)\in S_{T}$ and for all simple edges $(u_{1},u),...,(u_{n},u)\in E_Q$, there exists 
    \emph{\textbf{at least $n$ distinct parents}} $v_{1},...,v_{n}$ of $v$ in $G$ such that: $(u_{1}, v_{1}),...,(u_{n}, v_{n})\in S_{T}$.
    \item For each $(u,v)\in S_{T}$ and for each edge $e=(u, u^{'})$ in $E_Q$ with $\mathcal{C}(e)$=``$\geq p$'', there exists 
    \emph{\textbf{at least $p$ distinct children}} $v_{1},...,v_{p}$ of $v$ in $G$ such that: $(u^{'}, v_{1}),...,(u^{'}, v_{p})\in S_{T}$.
\end{enumerate}
\noindent $\mathcal{M}_{T}(Q,G)$ is the match result that corresponds to the maximum match relation $S_{T}$\footnote{Each subgraph in this match result satisfies 
\emph{CPL} relationships as well as \emph{CQs$^{+}$} of $Q$.}.
\end{my_definition}

Intuitively, we enhance the old definition of triple simulation with the condition (5) in order to preserve \emph{CQs$^{+}$} of $Q$. This condition 
requires that, for each edge $u\xrightarrow{\geq p}u^{'}$ in $Q$, each match $v$ of $u$ in $G$ must have at least $p$ distinct children that match 
all the child $u^{'}$ of $u$.

\begin{figure}[t]
\rule{\linewidth}{0.5pt}
\small
\begin{flushleft}
\textbf{Procedure} \textsc{\textbf{LR\_Checking}}($Q, G, u, v$)\newline
\textit{Input}: A QGP $Q(V_Q,E_Q,\lambda_Q,\mathcal{C})$ with only \emph{CQs$^{+}$}, a data graph $G(V,E,\lambda)$, a node $u\in V_Q$ with a 
potential match $v\in V$.\newline
\textit{Output}: Whether \emph{LR} constraints and \textit{CQs$^{+}$} defined over $u$ are satisfied by children and/or parents of $v$.
\end{flushleft}
\begin{algorithmic}[1]
\State \textit{\textbf{BG$_1$}} := $(X_1 \cup Y_1,E_1)$; \textit{\textbf{BG$_2$}} := $(X_2 \cup Y_2,E_2)$; where $X_1=Y_1=X_2=Y_2=E_1=E_2=\emptyset$;
  \ForAll{\textbf{child} $u^{'}$ of $u$ \textbf{with} $u^{'}\in \textbf{\textsc{LR}}(Q,u)$}
      \State $X_1$ := $X_1$ $\cup$ \{$u^{'}$\};
      \State \textbf{Do} $Y_1$ := $Y_1$ $\cup$ \{$v^{'}$\}; $E_1$ := $E_1$ $\cup$ \{($u^{'},v^{'}$)\}; \textbf{for each} ($v^{'}\in$ \textsc{sim}($u^{'}$) \textbf{with} $(v,v^{'})\in E$); 
  \EndFor
  
  \Statex $~~~~~~~~~~~~$\textbf{/*} \textsc{Consider the \textit{CQs$^{+}$} defined over the children of $u$} \textbf{*/}
  \ForAll{\textbf{child} $u^{'}$ of $u$ \textbf{with} $\mathcal{C}(u,u^{'})$=``$\geq p$'' \textbf{and} $p > 1$}
      \State $X_1$ := $X_1$ $\cup$ \{$u^{'}_1,\dots,u^{'}_p$\};\hfill\textbf{/*}\textsc{Create $p$ copies of the child $u^{'}$}\textbf{*/}
      \ForAll{$v^{'}\in$ \textsc{sim}($u^{'}$) \textbf{with} $(v,v^{'})\in E$}
	  \State $Y_1$ := $Y_1$ $\cup$ \{$v^{'}$\};
	  \ForAll{\textbf{copy} $u^{'}_i$ of $u^{'}$ in $X_1$}
	       \State $E_1$ := $E_1$ $\cup$ \{($u^{'}_i,v^{'}$)\};
	  \EndFor
      \EndFor
  \EndFor
  
  \ForAll{\textbf{parent} $u^{'}$ of $u$ \textbf{with} $u^{'}\in \textbf{\textsc{LR}}(Q,u)$}
      \State $X_2$ := $X_2$ $\cup$ \{$u^{'}$\};
      \State \textbf{Do} $Y_2$ := $Y_2$ $\cup$ \{$v^{'}$\}; $E_2$ := $E_2$ $\cup$ \{($u^{'},v^{'}$)\}; \textbf{for each} ($v^{'}\in$ \textsc{sim}($u^{'}$) \textbf{with} $(v^{'},v)\in E$);
  \EndFor
  
  \State \Return $true$ \textbf{if} (\textit{\textbf{CompleteMatch}}(\textit{\textbf{BG$_1$}}) \textbf{\&} \textit{\textbf{CompleteMatch}}(\textit{\textbf{BG$_2$}})); 
  \textbf{and $false$ otherwise};
\normalsize
\end{algorithmic}
\rule{\linewidth}{0.5pt}
\caption{New version of procedure \textbf{LR\_Checking} to handle \textit{CQs$^{+}$}.}
\label{LO_Prop_Checking_Procedure_Extended}
\end{figure}

\subsection{Adapting \textsc{TSim} for CQs$^{+}$}
Given a quantified pattern graph $Q$($V,E,\lambda,\mathcal{C}$) where $\mathcal{C}$ represents only \emph{CQs$^{+}$}. 
A new definition of the procedure \textbf{LR\_Checking} is given in Fig. \ref{LO_Prop_Checking_Procedure_Extended} in order 
to handle \textit{CQs$^{+}$}. 
Given a query node $u$ with a potential match $v$. As explained above, we construct two bipartite graphs $BG_1$ and $BG_2$ that inspect 
the \emph{LR} constraints defined over children and parents of $u$ respectively. 
Recall that \textit{CQs$^{+}$} are defined over children only. 
Thus, the equivalent \emph{LR} constraint of each one is defined and included in $BG_1$ [lines 6-14]. 
For each child $u^{'}$ of $u$ that is concerned by a \textit{CQ$^{+}$} of cardinality $p$ [line 6], we create $p$ copies of $u^{'}$ in $X_1$ 
[line 7]. Each potential match of $u^{'}$ is also a potential match of each copy of $u^{'}$. For this reason, 1) we add into $Y_1$ each child 
$v^{'}$ of $v$ that matches the child $u^{'}$ of $u$; and 2) we create an edge between each copy $u^{'}_i$ and $v^{'}$ to say that this copy can be 
matched by $v^{'}$. The resulting bipartite graph $BG_1$ inspects: 1) the \emph{LR} constraints defined over children of $u$; and 2) each \emph{LR} 
constraint that results from the transformation of a \textit{CQ$^{+}$} defined over some child of $u$. If a complete matching exists over $BG_1$ 
then all these \emph{LR} constraints are satisfied by children of $u$, i.e. all \textit{CQs$^{+}$} defined over children of $u$ are also 
satisfied (Proposition \ref{proposition:Translate_CQs_Into_LO_Prop}).

\begin{figure}[t]
\centering
   \includegraphics[width=300px,height=120px]{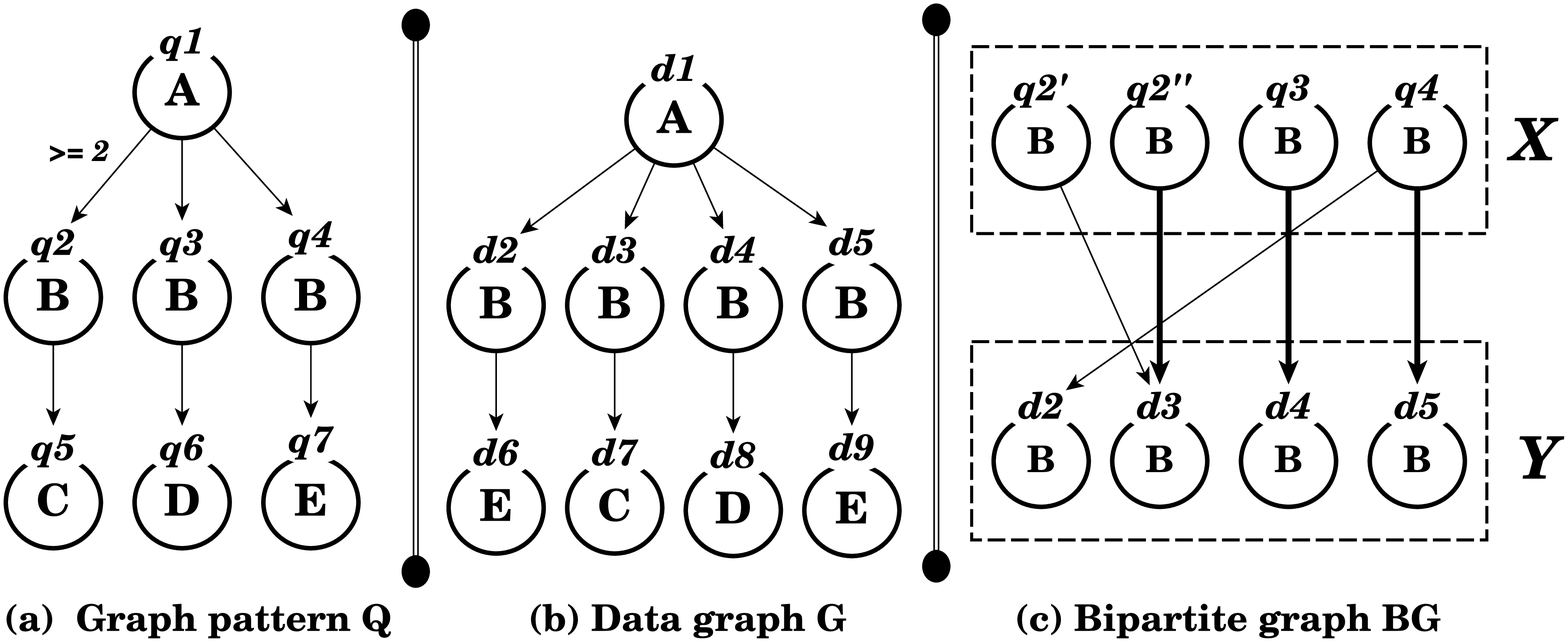}
   \caption{Satisfy \emph{CQs$^{+}$} as \emph{LR} constraints.}
   \label{Satisfy_CQs_As_LO_Prop}
\end{figure}

\begin{example}
Consider the quantified pattern graph $Q$ and the data graph $G$ depicted in side (a) and (b) of Fig. \ref{Satisfy_CQs_As_LO_Prop} respectively. 
It is clear to see that the \emph{LR} constraint, defined over the children $q_3$ and $q_4$ of $q_1$, is satisfied over $G$: 
by matching $q_1$, $q_3$, $q_4$ with $d_1$, $d_4$, and $d_5$ respectively. 
However, the \textit{CQ$^{+}$} ``$q_1\xrightarrow{\geq 2} q_2$'' is not satisfied. 
The match $d_1$ of $q_1$ must have at least two child nodes such that: each one is labeled with $B$ and have a child node labeled with 
$C$. Consider the couple $(q_1,d_1)$, the bipartite graph $BG_1$ constructed by the new procedure \textbf{LR\_Checking} is given 
in Fig. \ref{Satisfy_CQs_As_LO_Prop} (c). See that two copies of $q_2$ are created ($q_2^{'}$ and $q_2^{''}$) and each one is connected 
to the unique match $d_3$ of $q_2$. Since there is no complete matching over $BG_1$, the procedure returns $false$ which means that 
the \emph{LR} constraint and the \textit{CQ$^{+}$}, that are defined over children of $q_1$, are not all satisfied by children of $d_1$.
\end{example}

The next result states that the problem of matching pattern graphs with numeric aggregates is in PTIME when it is treated as an 
extension of graph simulation, contrary to the NP-Completeness found in \cite{Counting_Quantifiers_Fan_16} when the problem is 
studied under subgraph isomorphism.

\begin{theorem}\label{theorem:Overall_Time_Complexity_TSim_With_CQs}
Given a data graph $G$($V,E$) and a quantified pattern graph $Q$($V_Q,E_Q,\lambda_Q,\mathcal{C}$) where $\mathcal{C}$ defines only 
\textit{CQs$^{+}$}. Let $p$ be the largest cardinality of numeric aggregates $Q$. Algorithm \textbf{\textsc{TSim}} takes 
at most $O(|Q||G|+p.|V_Q|^{3}|V|^{2}\sqrt{p.|V_Q|+|V|})$ time 
to decide whether $G\prec_T Q$ and to find the match result $\mathcal{M}_{T}(Q,G)$.
\end{theorem}

Here algorithm \textbf{TSim} uses the new version of procedure \textbf{LR\_Checking} given in Fig. \ref{LO_Prop_Checking_Procedure_Extended}. 
The overall time complexity of algorithm \textbf{TSim}, in case of pattern graphs with numeric aggregates, is bounded by $O(p^{1.5}|Q|^{2}|G|^{1.5})$ 
where $p$ is bounded as follows: $1\leq p\leq |V|$.
\end{document}